\documentclass[prc,preprintnumbers,onecolumn]{revtex4}
\usepackage{graphicx}
\usepackage{amssymb}
\usepackage{amsmath}

\begin{document}

\title{Sub-barrier capture reactions with $^{16,18}$O and $^{40,48}$Ca beams}
\author{V.V.Sargsyan$^1$, G.G.Adamian$^{1}$, N.V.Antonenko$^1$, W. Scheid$^2$, and  H.Q.Zhang$^3$
}
\affiliation{$^{1}$Joint Institute for Nuclear Research, 141980 Dubna, Russia\\
$^{2}$Institut f\"ur Theoretische Physik der
Justus--Liebig--Universit\"at,
D--35392 Giessen, Germany\\
$^{3}$China Institute of Atomic Energy, Post Office Box 275, Beijing 102413,  China
}
\date{\today}

\begin{abstract}
Various   sub-barrier capture  reactions with beams $^{16,18}$O and $^{40,48}$Ca
are treated   within the quantum diffusion approach. The role of neutron transfer in these capture
reactions is discussed. The quasielastic and capture barrier distributions are analyzed and
compared with the recent experimental data.

\end{abstract}


 \maketitle

\section{Introduction}
From the present experimental data the role of the neutron
transfer channel in the capture (fusion) process cannot be
unambiguously inferred~\cite{Bertulani,Jia,Kol,EPJSub,EPJSub1}.
The fusion excitation functions have been recently measured for the reactions
$^{16}$O+$^{76}$Ge and $^{18}$O+$^{74}$Ge at energies near and below the Coulomb barrier
and the fusion barrier distributions have been extracted from the corresponding
excitation functions~\cite{Jia}. The fusion enhancement due to the positive $Q_{2n}$-value
two neutron ($2n$) transfer channel for $^{18}$O+$^{74}$Ge has not been revealed as compared with the reference
system $^{16}$O+$^{76}$Ge~\cite{Jia}.
This is very different from the situation for the reactions  $^{40}$Ca+$^{124,132}$Sn~\cite{Kol} and other 
systems in literature, which show considerable sub-barrier enhancements.	
The enhancement appears to be related to the existence of  positive $Q$ values for neutron transfer.

The purpose of this paper is the theoretical explanation of these experimental observations.
Within the quantum diffusion approach~\cite{EPJSub,EPJSub1}
we try to  answer  the question how strong
the influence of neutron transfer  in sub-barrier capture (fusion) reactions
$^{18}$O+$^{74}$Ge,$^{52,50}$Cr,$^{94,92}$Mo,$^{112,114,118,120,124,126}$Sn and $^{40,48}$Ca+$^{124,132}$Sn.
This study seems to be important for future  experiments indicated in Ref.~\cite{Jia}.
In addition, the new structures of the quasielastic and capture barrier
distributions at deep sub-barrier
energies will be discussed.

\section{Model}
In the quantum diffusion approach~\cite{EPJSub,EPJSub1}
the collisions of  nuclei are described with
a single relevant collective variable: the relative distance  between
the colliding nuclei. This approach takes into consideration the fluctuation and dissipation effects in
collisions of heavy ions which model the coupling with various channels
(for example, coupling of the relative motion with low-lying collective modes
such as dynamical quadrupole and octupole modes of the target and projectile~\cite{Ayik333}).
We have to mention that many quantum-mechanical and non-Markovian effects accompanying
the passage through the potential barrier are taken into consideration in our
formalism~\cite{EPJSub1}.
The  nuclear deformation effects
are taken into account through the dependence of the nucleus-nucleus potential
on the deformations and mutual orientations of the colliding nuclei.
To calculate the nucleus-nucleus interaction potential $V(R)$,
we use the procedure presented in Refs.~\cite{EPJSub1}.
For the nuclear part of the nucleus-nucleus
potential, the double-folding formalism with
the Skyrme-type density-dependent effective
nucleon-nucleon interaction is used.
With this approach many heavy-ion capture
reactions at energies above and well below the Coulomb barrier have been
successfully described~\cite{EPJSub1}.
%
Note that the diffusion models, which  include the quantum statistical effects,
were also treated in Refs.~\cite{Hofman}.

Following the hypothesis of Ref.~\cite{Broglia},
we assume that the sub-barrier capture
in the reactions under consideration
mainly  depends  on the possible two-neutron
transfer with the  positive  $Q_{2n}$-value.
Our assumption is that, just before the projectile is captured by the target-nucleus
(just before the crossing of the Coulomb barrier) which is a slow process,
the  $2n$-transfer ($Q_{2n}>0$) occurs   that can lead to the
population of the
excited collective
states in the recipient nucleus~\cite{SSzilner}.
So, the motion to the
$N/Z$ equilibrium starts in the system before the capture because it is energetically favorable
in the dinuclear system in the vicinity of the Coulomb barrier.
For the reactions  considered,
the average change of mass asymmetry is related to the two-neutron
transfer. In  these reactions  the
$2n$-transfer channel is more favorable than $1n$-transfer channel ($Q_{2n}>Q_{1n}$).
Since after the $2n$-transfer the mass numbers,  the deformation parameters
of the interacting nuclei, and, correspondingly, the height $V_b=V(R_b)$
of the Coulomb barrier are changed,
one can expect an enhancement or suppression of the capture.
This scenario was verified in the description of many reactions~\cite{EPJSub1}.

\section{Results of calculations}
All calculated results are obtained with the same set of parameters
 as in Ref.~\cite{EPJSub1}
and are rather insensitive
to the reasonable variation of them~\cite{EPJSub1}.
Realistic friction coefficient in the momentum
$\hbar\lambda$=2 MeV
is
used which
is close to those calculated within the mean field approaches~\cite{obzor}.
The parameters of the nucleus-nucleus interaction potential $V(R)$
are adjusted to describe the experimental
data at energies above the Coulomb barrier corresponding to spherical nuclei.
The absolute values of the quadrupole deformation parameters $\beta_2$
of even-even deformed nuclei are taken from Ref.~\cite{Ram}.
In Ref.~\cite{Ram}, the quadrupole
deformation parameters $\beta_2$ are given for the first excited
2$^{+}$ states of nuclei. For the  nuclei deformed in the
ground state, the $\beta_2$ in 2$^{+}$ state is similar
to the $\beta_2$ in the ground state and we use $\beta_2$
from Ref.~\cite{Ram} in the calculations.
For the double magic   nucleus $^{16}$O,
in the ground state
we take $\beta_2=0$.
Since there are  uncertainties in the definition of the values of $\beta_2$
in  light- and medium-mass nuclei,
one can extract the
quadrupole deformation parameters of
these  nuclei from a comparison
of the calculated capture cross sections with the existing experimental data.
By describing the reactions
$^{18}$O+$^{208}$Pb,
where there are no neutron transfer channels with positive $Q$-values,
we extract $\beta_2=0.1$ for the ground-state
of  $^{18}$O~\cite{EPJSub1}.
This extracted value is used in our calculations.

\subsection{Effect of neutron transfer in reactions with beams $^{40,48}$Ca}
To eliminate the influence
of the nucleus-nucleus
potential on the capture (fusion) cross section and to make conclusions about the role of deformation of
colliding nuclei and the nucleon transfer between interacting nuclei in the capture  (fusion) cross
section,  a reduction procedure is useful~\cite{Gomes}. 
It consists of the following transformations:
$$E_{\rm c.m.} \rightarrow x= \dfrac{E_{\rm c.m.}-V_b}{\hbar \omega_b},
\qquad \sigma_{cap} \rightarrow \sigma_{cap}^{red}=\dfrac{2 E_{\rm c.m.}}{\hbar \omega_b R_b^{2}}\sigma_{cap},$$
where  $\sigma_{cap}=\sigma_{cap}(E_{\rm c.m.})$ is the capture cross section at bombarding energy 
$E_{\rm c.m.}$.
The frequency $\omega_b =\sqrt{V^{''}(R_b)/\mu}$
is related with the second derivative $V^{''}(R_b)$ of the total nucleus-nucleus potential $V(R)$
(the Coulomb + nuclear parts)
at the barrier radius $R_b$ and the reduced mass parameter $\mu$.
With these replacements we  compared  the  reduced calculated capture (fusion) cross sections
$\sigma_{cap}^{red}$ for the reactions $^{40,48}$Ca+$^{124,132}$Sn (Fig.~1).
The choice of the projectile-target combination is crucial, and for the systems
studied one can  make unambiguous statements
regarding the neutron transfer process with a positive $Q$-value when the interacting nuclei
are  double magic or semi-magic spherical nuclei.
In this case one can disregard the strong  direct nuclear deformation effects. 
In Fig.~1,
one can see that the reduced capture cross sections in the  reactions $^{40}$Ca+$^{124,132}$Sn
with the positive $Q_{2n}$-values
strongly deviate from   those in the reactions
$^{48}$Ca+$^{124,132}$Sn,
where the neutron transfers   are suppressed because of the negative $Q$-values.
\begin{figure}
\includegraphics[scale=1]{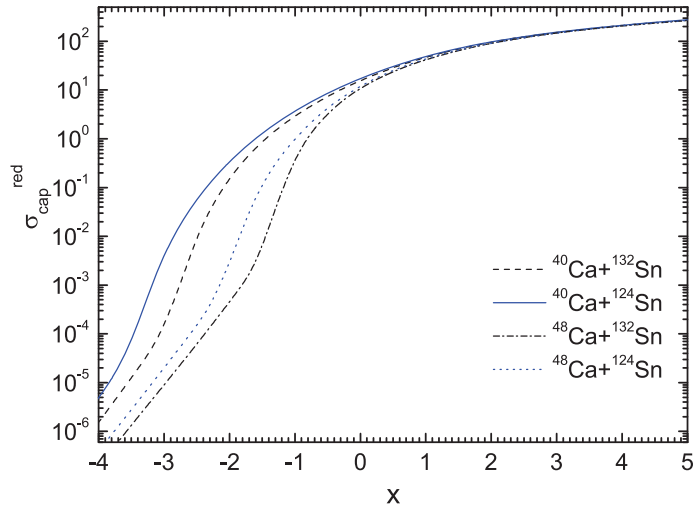}
\caption{(Colour online) The calculated reduced capture cross sections versus $(E_{\rm c.m.}-V_b)/(\hbar\omega_b)$ 
in the reactions
$^{40}$Ca+$^{124}$Sn (solid line),
$^{48}$Ca+$^{124}$Sn (dashed line), $^{48}$Ca+$^{124}$Sn (dotted line),
and  $^{48}$Ca+$^{132}$Sn (dash-dotted line).
}
\label{1_fig}
\end{figure}
 After two-neutron transfer in the reactions
 $^{40}$Ca($\beta_2=0$)+$^{124}$Sn($\beta_2=0.1$)$\to ^{42}$Ca($\beta_2=0.25$)+$^{122}$Sn($\beta_2=0.1$) 
 ($Q_{2n}$=5.4 MeV) 
 and
$^{40}$Ca($\beta_2=0$)+$^{132}$Sn($\beta_2=0$)$\to ^{42}$Ca($\beta_2=0.25$)+$^{130}$Sn($\beta_2=0$) 
($Q_{2n}$=7.3 MeV)
the deformation of the light nucleus increases and the mass asymmetry of the system decreases  and,
thus, the value of the Coulomb barrier decreases and
the capture cross section becomes larger (Fig.~1). So, because of the transfer effect 
the systems $^{40}$Ca+$^{124,132}$Sn show large sub-barrier enhancements with respect to 
the systems $^{48}$Ca+$^{124,132}$Sn. 
We observe that the $\sigma_{cap}^{red}$ in the  $^{40}$Ca+$^{124}$Sn ($^{48}$Ca+$^{124}$Sn)   
reaction are larger
than those in the  $^{40}$Ca+$^{132}$Sn  ($^{48}$Ca+$^{132}$Sn)   reaction. The reason of that
is the nonzero quadrupole deformation  of the heavy nucleus $^{124}$Sn. 
It should be stressed that there are almost no difference 
between  $\sigma_{cap}^{red}$ in the reactions $^{40,48}$Ca+$^{124,132}$Sn
 at energies above the Coulomb barrier. 

\begin{figure}
\includegraphics[scale=1]{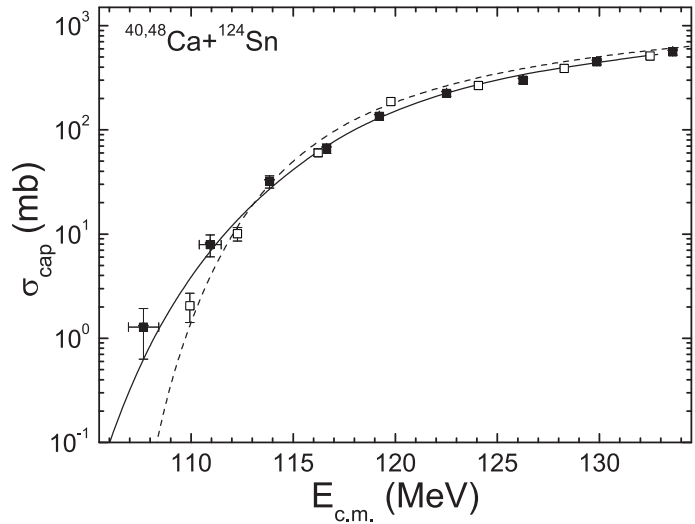}
\caption{The calculated  capture cross sections versus $E_{\rm c.m.}$ for the reactions
$^{40}$Ca+$^{124}$Sn (solid line) and $^{48}$Ca+$^{124}$Sn (dashed line).
The experimental data  for the reactions $^{40}$Ca+$^{124}$Sn (solid squares) 
and $^{48}$Ca+$^{124}$Sn (open squares) are from Ref.~\protect\cite{Kol}.
In the calculations the barriers were adjusted to the experimental values.
}
\label{2_fig}
\end{figure}
\begin{figure}
\includegraphics[scale=1]{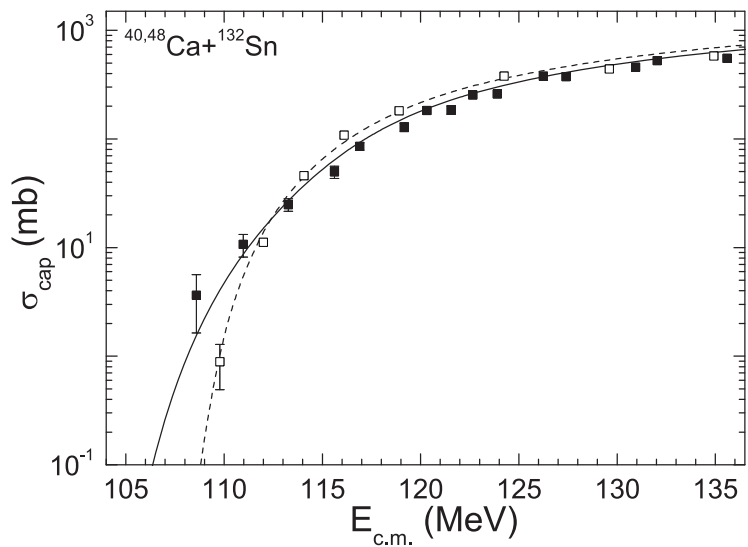}
\caption{The calculated  capture cross sections versus $E_{\rm c.m.}$ for the reactions
$^{40}$Ca+$^{132}$Sn (solid line) and $^{48}$Ca+$^{132}$Sn (dashed line).
The experimental data  for the reactions $^{40}$Ca+$^{132}$Sn (solid squares) and $^{48}$Ca+$^{132}$Sn (open squares)
are from Ref.~\protect\cite{Kol}.
In the calculations the barriers were adjusted to the experimental values.
}
\label{3_fig}
\end{figure}
In Figs.~2 and 3 one can see a good agreement between the  calculated results and the experimental data 
in the reactions $^{40,48}$Ca+$^{124,132}$Sn.
This means that the observed capture enhancements  in
the reactions $^{40}$Ca+$^{124,132}$Sn   at sub-barrier energies 
are related to the two-neutron transfer effect.
Note that the slope of the excitation function strongly depends on the deformations of the 
interacting nuclei and, respectively, on the neutron transfer effect.

To describe the reactions $^{40,48}$Ca+$^{132}$Sn (Fig.~2) and  $^{48}$Ca+$^{124,132}$Sn (Fig.~3),
we extracted the values of the corresponding  Coulomb barrier $V_b$ for the spherical nuclei.
There are  differences between  the calculated and extracted $V_b$. 
From the direct  calculations 
of the nucleus-nucleus potentials (with the same set of parameters), we obtained
$V_b$($^{40}$Ca+$^{124}$Sn)-$V_b$($^{48}$Ca+$^{124}$Sn)=2.3 MeV,
$V_b$($^{40}$Ca+$^{132}$Sn)-$V_b$($^{48}$Ca+$^{132}$Sn)=2.2 MeV, 
$V_b$($^{40}$Ca+$^{124}$Sn)-$V_b$($^{40}$Ca+$^{132}$Sn)=1.3 MeV, 
and
$V_b$($^{48}$Ca+$^{124}$Sn)-$V_b$($^{48}$Ca+$^{132}$Sn)=1.2 MeV. 
From the extractions, we got 
$V_b$($^{40}$Ca+$^{124}$Sn)-$V_b$($^{48}$Ca+$^{124}$Sn)=1.1 MeV
$V_b$($^{40}$Ca+$^{132}$Sn)-$V_b$($^{48}$Ca+$^{132}$Sn)=1.0 MeV, 
$V_b$($^{40}$Ca+$^{124}$Sn)-$V_b$($^{40}$Ca+$^{132}$Sn)=-0.3 MeV,   
and
$V_b$($^{48}$Ca+$^{124}$Sn)-$V_b$($^{48}$Ca+$^{132}$Sn)=-0.4 MeV, 
which seem to be  unrealistically small. 
However,  these differences of $V_b$ 
do not influence the slopes of the excitation functions but only 
lead to the shifting of the energy scale.
With realistic isospin trend of $V_b$   
$\sigma_{cap}$($^{40}$Ca+$^{124}$Sn)$< \sigma_{cap}$($^{48}$Ca+$^{124}$Sn) 
and
$\sigma_{cap}$($^{40}$Ca+$^{132}$Sn)$< \sigma_{cap}$($^{48}$Ca+$^{132}$Sn) 
at energies above the corresponding Coulomb barriers.

\subsection{Effect of neutron transfer in reactions with beams $^{16,18}$O}
Figures  4-7 show the capture excitation function  for the reactions
$^{16,18}$O+$^{76,74}$Ge,
$^{16,18}$O+$^{94,92}$Mo,
$^{16,18}$O+$^{114,112,120,118,126,124}$Sn,  and $^{16,18}$O+$^{52,50}$Cr
as a function of  bombarding energy.
One can see a rather good agreement between the
calculated results and the experimental data~\cite{Jia,16OAGe,AO92Mo,AOASn}
for the reactions
$^{16}$O+$^{76}$Ge,
$^{16,18}$O+$^{92}$Mo, and
$^{18}$O+$^{112,118,124}$Sn.
\begin{figure}
\includegraphics[scale=1]{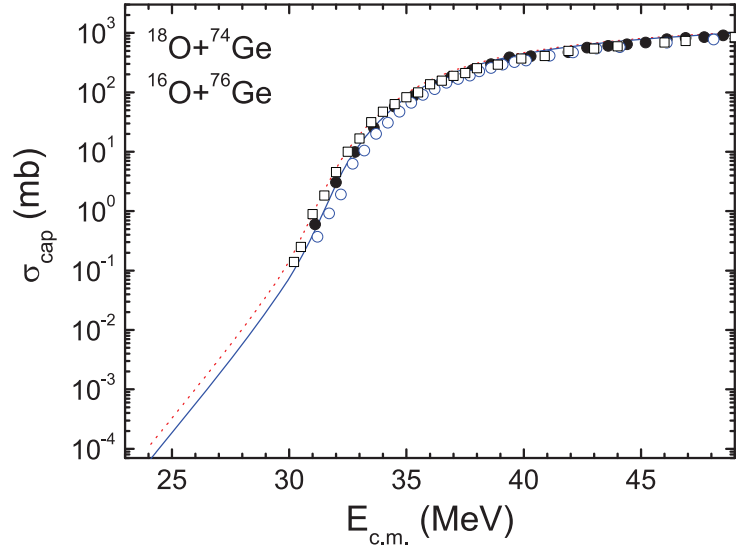}
\caption{(Colour online) 
The calculated (solid line) capture cross sections versus $E_{\rm c.m.}$ for the reactions
$^{16}$O+$^{76}$Ge  and $^{18}$O+$^{74}$Ge (the curves coincide).
For the $^{18}$O+$^{74}$Ge reaction,
the calculated capture cross sections without
 neutron transfer are shown by dotted line.
The experimental data  for the reactions $^{16}$O+$^{76}$Ge (open circles) and $^{18}$O+$^{74}$Ge (open squares)
are from Ref.~\protect\cite{Jia}.
The experimental data  for the $^{16}$O+$^{76}$Ge reaction (solid circles)
are from Ref.~\protect\cite{16OAGe}.
}
\label{4_fig}
\end{figure}
\begin{figure}
\includegraphics[scale=1]{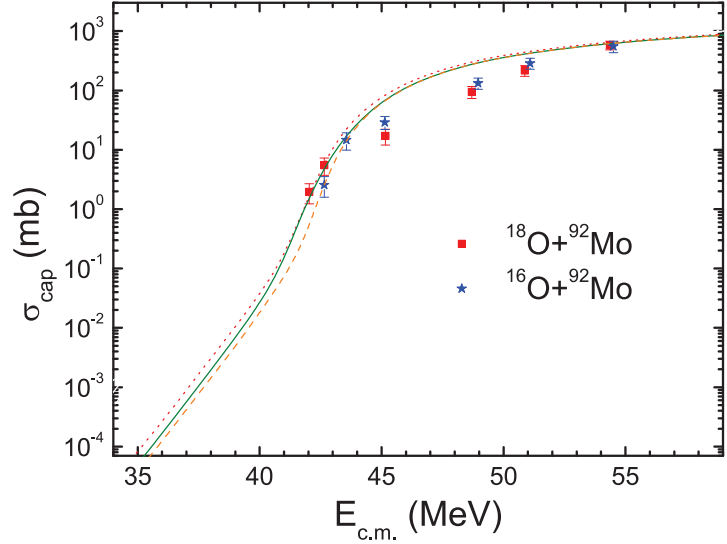}
\caption{(Colour online) The calculated  capture cross sections versus $E_{\rm c.m.}$ for the reactions
$^{16}$O+$^{92}$Mo (dashed line) and $^{18}$O+$^{92}$Mo (solid line).
For the $^{18}$O+$^{92}$Mo reaction,
the calculated capture cross sections without
the neutron transfer are shown by dotted line.
The experimental data  for the reactions
$^{16}$O+$^{92}$Mo (solid stars) and $^{18}$O+$^{92}$Mo (solid squares)
are from Ref.~\protect\cite{AO92Mo}.
}
\label{5_fig}
\end{figure}
\begin{figure}
\includegraphics[scale=1.4]{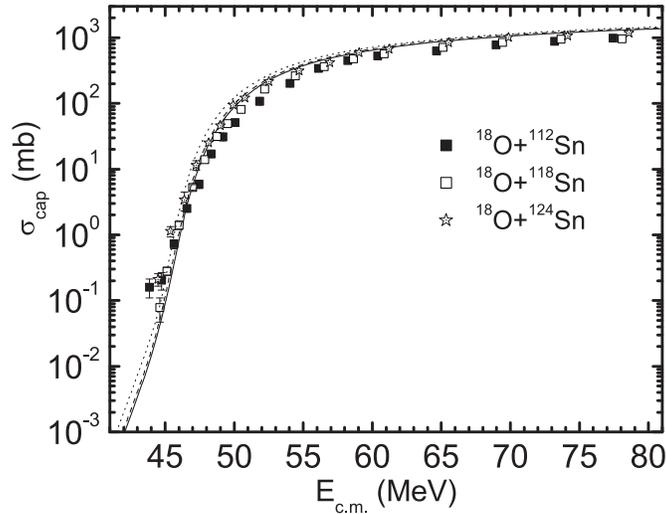}
\caption{The calculated  capture cross sections versus $E_{\rm c.m.}$ for the reactions
$^{16}$O+$^{114}$Sn  and $^{18}$O+$^{112}$Sn (solid line),
$^{16}$O+$^{120}$Sn and
$^{18}$O+$^{118}$Sn (dashed line), $^{16}$O+$^{126}$Sn
and $^{18}$O+$^{124}$Sn (dotted line).
The calculated results
for the reactions
$^{16}$O+$^{114,120,126}$Sn and $^{18}$O+$^{112,118,124}$Sn  coincide, respectively.
The experimental data  for the  reactions $^{18}$O+$^{112}$Sn (solid squares), $^{18}$O+$^{118}$Sn (open squares),
and $^{18}$O+$^{124}$Sn (open stars)
are from Ref.~\protect\cite{AOASn}.
}
\label{6_fig}
\end{figure}
\begin{figure}
\includegraphics[scale=1]{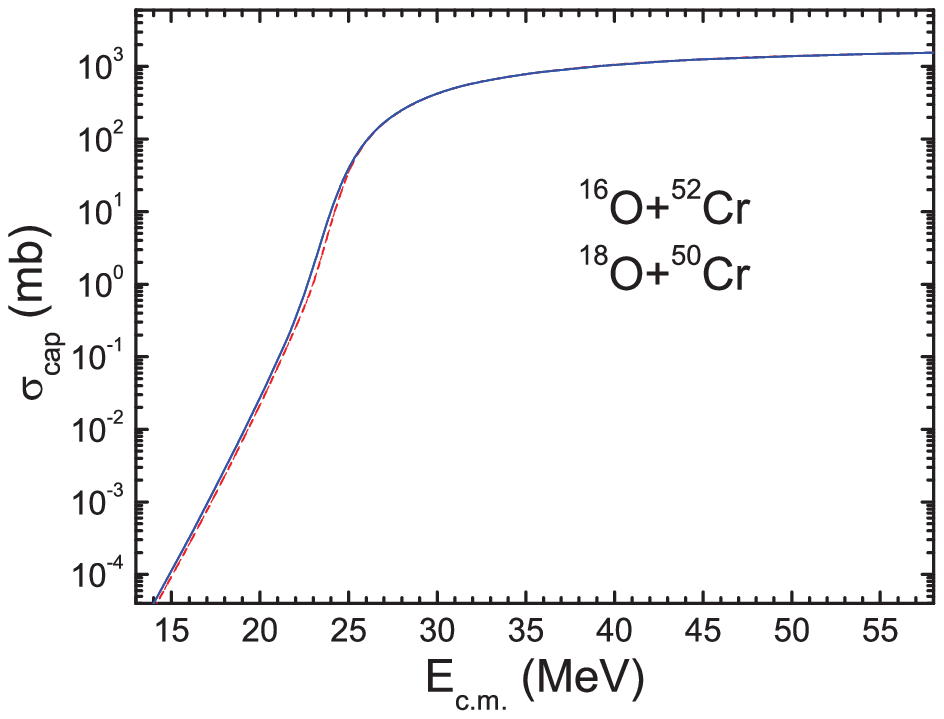}
\caption{(Coulor online) The calculated  capture cross sections versus $E_{\rm c.m.}$ for the reactions
$^{16}$O+$^{52}$Cr (dashed line) and $^{18}$O+$^{50}$Cr (solid line).
}
\label{7_fig}
\end{figure}
The $Q_{2n}$-values for the $2n$-transfer
processes are positive (negative) for all reactions with $^{18}$O ($^{16}$O).
Thus, the neutron transfer can be important for the reactions with  the $^{16}$O beam.
However, our results show that cross sections for reactions $^{16}$O+$^{76}$Ge ($^{16}$O+$^{114,120,126}$Sn,$^{52}$Cr)
and $^{18}$O+$^{74}$Ge ($^{18}$O+$^{114,118,124}$Sn,$^{50}$Cr) are very similar.
The reason of such behavior is that after the $2n$-transfer in the system $^{18}$O+$^{A-2}$X$\to ^{16}$O+$^{A}$X the deformations
remain to be  similar.
As a result, the corresponding Coulomb barriers of the  systems
$^{18}$O+$^{A-2}$X and $^{16}$O+$^{A}$X
 are almost the same and, correspondingly, their capture cross
sections  coincide.
Just the same behavior
was observed in the recent experiments $^{16,18}$O+$^{76,74}$Ge~\cite{Jia}.

One can see in Figs. 4-7 that at energies above and near  the Coulomb barrier
the cross sections with and without
two-neutron transfer are quite similar.
After the $2n$-transfer (before the capture)
in the reactions
$^{18}$O($\beta_2=0.1$) + $^{92}$Mo($\beta_2=0.05$)$\to ^{16}$O($\beta_2=0$) + $^{94}$Mo($\beta_2=0.151$),
$^{18}$O($\beta_2=0.1$) + $^{74}$Ge($\beta_2=0.283$)$\to ^{16}$O($\beta_2=0$) + $^{76}$Ge($\beta_2=0.262$),
$^{18}$O($\beta_2=0.1$)+$^{112}$Sn($\beta_2=0.123$)$\to ^{16}$O($\beta_2=0$)+$^{114}$Sn($\beta_2=0.121$),
$^{18}$O($\beta_2=0.1$)+$^{118}$Sn($\beta_2=0.111$)$\to ^{16}$O($\beta_2=0$)+$^{120}$Sn($\beta_2=0.104$),
and
$^{18}$O($\beta_2=0.1$)+$^{124}$Sn($\beta_2=0.095$)$\to ^{16}$O($\beta_2=0$)+$^{126}$Sn($\beta_2=0.09$)
the deformations of the nuclei  decrease  and
the values of the corresponding Coulomb barriers  increase.
As a result, the transfer
suppresses the capture process at the sub-barrier energies.
The suppression becomes  stronger with decreasing  energy.
As examples, in Fig.~4 and 5 we show this effect for the reactions
$^{18}$O+$^{74}$Ge,$^{92}$Mo.

\subsection{Capture and quasielastic barrier distributions}

In Figs.~8 and 9, the calculated capture  barrier distributions
$$D=d^2(E_{\rm c.m.}\sigma_{cap})/dE_{\rm c.m.}^2$$
for the reactions $^{16}$O+$^{76}$Ge,$^{144,154}$Sm
have only one pronounced maximum
around $E_{\rm c.m.}=V_b$ as in the experiments~\cite{Jia,Timmers}.
The calculated barrier distributions in Figs.~8 and 9 are slightly wider and  fit the experimental data
better than those obtained with the couple-channels approach in Fig.~5 of Ref.~\cite{Jia}.
The capture (fusion) cross sections for the reactions $^{16}$O+$^{76}$Ge,$^{144,154}$Sm
were well described with the quantum diffusion model in Ref.~\cite{EPJSub1}.
With almost spherical (deformed) target-nucleus
we obtain a more narrow (wide) barrier distribution for the
$^{16}$O+$^{144}$Sm ($^{16}$O+$^{154}$Sm) reaction.

We compared the capture and the quasielastic barrier distributions for these reactions (Figs.~8 and 9).
%
%
\begin{figure}
\includegraphics[scale=1]{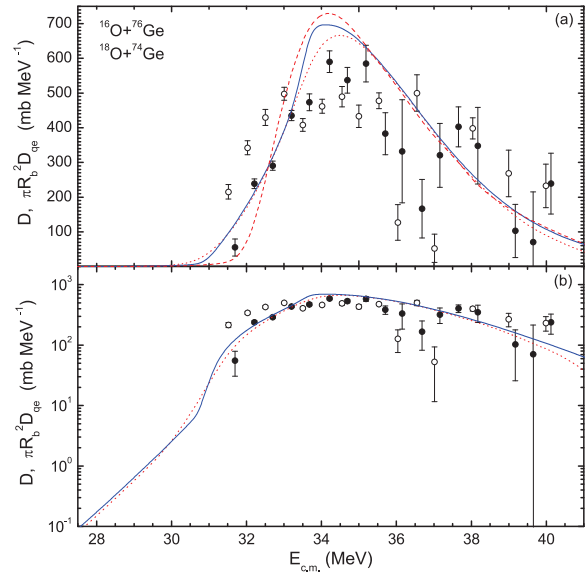}
\caption{(Colour online)
(a) The calculated values  of the quasielastic $\pi R_b^2D_{qe}$ (solid line) and capture $D$ (dotted line)
barrier distributions for the reactions $^{16}$O + $^{76}$Ge and $^{18}$O + $^{74}$Ge. The curves coincide
for these reactions.
The calculated $D$ for the spherical interacting nuclei
is shown by dashed line.
The experimental data for the reactions $^{16}$O + $^{76}$Ge (solid circles) and $^{18}$O + $^{74}$Ge
(open circles)  are from Ref.~\protect\cite{Jia}.
(b) The calculated values  of  $\pi R_b^2D_{qe}$  (solid line)
and $D$ (dotted line) are shown in the  logarithmic scale.
}
\label{8_fig}
\end{figure}
\begin{figure}
\includegraphics[scale=1]{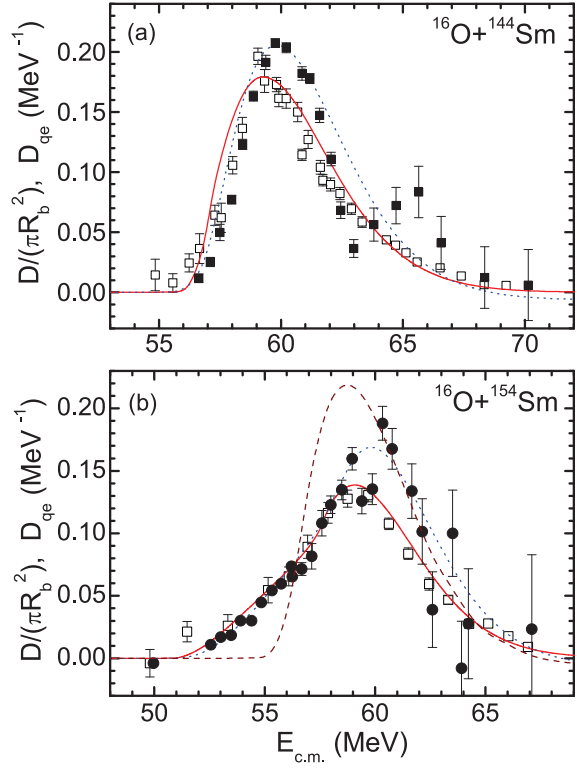}
\caption{(Colour online)
The calculated values of quasielastic $D_{qe}$ (solid line)
and capture $D/(\pi R_b^2)$ (dotted line) barrier distributions
for the reactions $^{16}$O + $^{144}$Sm (a) and $^{16}$O + $^{154}$Sm (b).
The experimental $D_{qe}$ (open squares) and  $D/(\pi R_b^2)$ (solid circles) are
from Ref.~\protect\cite{Timmers}. The calculated $D$ for the spherical interacting nuclei
is shown by dashed line (b).
}
\label{9_fig}
\end{figure}
There is a direct relationship between the capture
and the quasielastic scattering processes because any loss from
the quasielastic channel contributes directly to the capture (the conservation of the reaction flux):
$$P_{qe}(E_{\rm c.m.},J)+P_{cap}(E_{\rm c.m.},J)=1$$
and
$$dP_{cap}/dE_{\rm c.m.}=-dP_{qe}/dE_{\rm c.m.},$$
where
$P_{qe}$  is the reflection probability and
$P_{cap}$ is the capture (transmission) probability ($J$ is the partial wave).
The quasielastic barrier distribution is extracted
by taking the first derivative of the $P_{qe}(E_{\rm c.m.},J=0)$ or $P_{cap}(E_{\rm c.m.},J=0)$
with respect to $E_{\rm c.m.}$, that is,
$$D_{qe}(E_{\rm c.m.})=-dP_{qe}(E_{\rm c.m.},J=0)/dE_{\rm c.m.}=dP_{cap}(E_{\rm c.m.},J=0)/dE_{\rm c.m.}.$$
So, by employing the quantum diffusion approach and calculating $dP_{cap}(E_{\rm c.m.},J=0)/dE_{\rm c.m.}$,
one can obtain $D_{qe}(E_{\rm c.m.})$.
One can see in Figs.~8 and 9 that the shapes of the
quasielastic and capture barrier distributions are similar. The same conclusion
was experimentally obtained  for the $^{20}$Ne+$^{208}$Pb reaction in Ref.~\cite{Piasecki}.
As in the case of capture barrier distribution, one can show that
the width of the quasielastic barrier distribution  increases
with the deformation of the target-nucleus.
In addition to the mean peak position of the  $D_{qe}$ around
the barrier height, we observe the sharp change of the slope of  $D_{qe}$ or $D$  below
the Coulomb barrier  energy because of a change of the
regime of interaction (the external turning point leaves the region of the nuclear forces
and friction~\cite{EPJSub,EPJSub1})
in the deep sub-barrier capture process (Fig.~8(b)).

\section{Summary}

As shown with the quantum diffusion approach,
 the capture cross sections for the reactions
 $^{16}$O+$^{52}$Cr,$^{76}$Ge,$^{94}$Mo,$^{114,120,126}$Sn and
 $^{18}$O+$^{50}$Cr,$^{74}$Ge,$^{92}$Mo,$^{112,118,124}$Sn, respectively,  almost match.
The fusion enhancement due to the positive $Q_{2n}$-value
$2n$-transfer  for $^{18}$O+$^{74}$Ge has not been observed~\cite{Jia} because the deformations of nuclei
slightly decrease after the neutron transfer.
This is different from the situation for the reactions    $^{40}$Ca+$^{124,132}$Sn~\cite{Kol} 
with large positive $Q_{2n}$-values. 
The strong enhancements  have been observed~\cite{Kol} in these reactions at sub-barrier
energies  because the deformation  of light nucleus strongly
increases (the heavy nucleus is spherical before and after transfer) 
after the two-neutron transfer.

We found that the shapes of the quasielastic and capture barrier distributions are   similar.
The sharp change of the slope of
 the quasielastic or capture barrier distribution is predicted at deep sub-barrier energy.
This anomalous behavior of the  barrier distribution is expected to be the experimental indication
of a  change of the regime of interaction  in the sub-barrier capture.
One concludes that the quasielastic technique could be an  important tool in capture (fusion) research.

This work was supported by DFG, NSFC, and RFBR.
The IN2P3(France) - JINR(Dubna)
and Polish - JINR(Dubna)
Cooperation Programmes are gratefully acknowledged.
 H.Q.~Zhang is grateful to Chinese NSFC 
for the partial support.\\


\end{document}